\documentclass[a4paper]{spie}  

\usepackage{amsmath,amsfonts,amssymb}
\usepackage{graphicx}
\usepackage[colorlinks=true, allcolors=blue]{hyperref}

\title{The Near-Ultraviolet eXplorer (NUX): a ground-based wide-field near-UV telescope to search for near-UV transients}

\author[a]{Rudy Wijnands}
\author[b]{Steven Bloemen}
\author[a]{Rasjied Sloot}
\author[c]{Rik ter Horst}
\author[b]{Andr\'e Young}
\author[a]{Mattijs Bakker}
\author[b,d,e]{Paul Groot}
\author[b]{Paul Vreeswijk}
\affil[a]{Anton Pannekoek Institute for Astronomy, University of Amsterdam, PO Box 94249, 1090 GE Amsterdam, The Netherlands}
\affil[b]{Department of Astrophysics/IMAPP, Radboud University, PO Box 9010, 6500 GL Nijmegen, The Netherlands}
\affil[c]{NOVA Optical IR Instrumentation Group at ASTRON, Oude Hoogeveensedijk 4,  7991 PD Dwingeloo, The Netherlands}
\affil[d]{Department of Astronomy and Inter-University Institute for Data Intensive Astronomy, University of Cape Town, Private Bag X3, Rondebosch 7701, South Africa}
\affil[e]{South African Astronomical Observatory, PO Box 9, Observatory, Cape Town 7935, South Africa}

\authorinfo{Further author information: Send correspondence to Rudy Wijnands at r.a.d.wijnands@uva.nl}

\begin{document} 
\maketitle

\begin{abstract}
We present the Near-Ultraviolet eXplorer (NUX), which will consist out of 4 small (36 cm diameter) ground-based telescopes that are optimized for the shortest wavelengths that are detectable from Earth (i.e., the near-UV [NUV] wavelength range of 300-350 nm). Each telescope will have a field-of-view of $\sim$17 square degrees sampled at $\sim$2.6”/pixel, and will reach a NUV magnitude (AB) of 20 in 2.5 minutes exposures (in dark time). The goal of NUX is to improve our understanding of the physical processes that power fast (days) to very fast (hours) hot transients, such as shock-breakout and shock-cooling emission of supernovae and the electromagnetic counterparts of gravitational wave events. Each telescope will be an off-the-shelf 14” Celestron RASA telescope, retrofitted with NUV optics. We have already demonstrated that the normal Schmidt corrector of this telescope can be replaced by a custom made one consisting of NUV transparent glass. Currently, a prototype NUX telescope is being fully assembled to demonstrate the technical and scientific feasibility of the NUX concept. Site tests will be held (in 2025/2026) at La Silla, Chile, to determine the NUV characteristics of the atmosphere at this site.
\end{abstract}


\section{INTRODUCTION}
\label{section:introduction}

Time-domain astronomy has become an intense field of research over the last two decades with many observing facilities searching and finding numerous transients at nearly all electromagnetic (EM) wavelengths (from $\gamma$-ray to radio; see, e.g., [\citenum{2019Galax...7...28R,2019PASP..131a8002B,2016mks..confE..13F}]), or using gravitational waves (GWs; i.e., their EM counterparts \cite{2017PhRvL.119p1101A,2017ApJ...848L..12A}) and even neutrinos \cite{2018Sci...361.1378I}. Such transients directly probe extreme (astro-)physics. Ultraviolet (UV) follow-up observations play a very important role in understanding these extreme processes. However, the UV itself has not yet been utilised in great depth to discover transients despite that most of them show peak luminosity, at early times, in the UV. Currently, no UV dedicated facility exists that can repeatedly scan large fractions of the sky, although experiments are being built such as ULTRASAT and UVEX (see Section \ref{section:ultrasatuvex} for more details about these observatories).  In addition, very recently a project was started\cite{2022A&A...663A...5M} to utilise the UVOT aboard Swift to search for serendipitous UV transients despite its very small field-of-view (FOV). However, this project is limited in scope and mostly very common types of transients are found (i.e., variable stars, outbursting accreting white dwarfs, and a large range of different types of accreting supermassive black holes; see [\citenum{2022A&A...663A...5M}] for preliminary results). The reason that currently there are no UV transient search facilities is mainly because most of the UV is blocked by the Earth's atmosphere and large FOV UV instruments on board satellites are expensive. However, despite this, the near-UV band (NUV; here defined as radiation in the wavelength range 300-350 nm\footnote{In the literature, multiple, slightly different, definition of NUV are used although most of them fall within the wavelength range of 200-400 nm (although not always; e.g., the GALEX NUV band covers $\sim$180-280 nm).}) can still be observed from the Earth's surface, potentially allowing for viable UV transient research using a ground-based NUV telescope. Here, we present  our proposed Near-UV eXplorer (NUX) that will operate, from the ground, in the NUV band. NUX will be located at La Silla (managed by the European Southern Observatory; ESO) and will allow us to systematically probe the transient NUV sky with the focus on the fast (days) to very fast (hours) transients.  

\section{SCIENCE GOALS}
\label{section:goals}

Since the NUV has hardly been used to search for and to study transients, the potential for new and unexpected discoveries with NUX is high (i.e., new types of transients or new phenomena of known source types). Finding such new source types or new source behavior is one of the prime goals of NUX. Other prime targets that will be investigated using NUX are cosmic explosions and transiently accreting compact objects. 

\paragraph{Cosmic explosions:} Cosmic explosions are violent, one-off events at cosmological distances. Therefore, the energy output from them is enormous. Typically these events are related to the death of stars and include GW events, gamma-ray bursts, and supernova explosions. Although the broad physical picture about these events is usually well established, many details about the physical processes powering them remain unclear and NUX would provide very important data to push the models further. We expect the largest NUX impact on the EM counter parts of GW events (the so-called kilonovae), the shock-break out and shock-cooling of core-collapse supernovae (CCSNe), as well as the recently identified class of fast blue optical transients (FBOTs). The common characteristics of these transient events are that they are hot systems (thus typically bright in the [N]UV) and that they evolve over time scales of hours to a few days (often peaking in the [N]UV before reaching their maximum in the optical). These objects are prime targets for NUX because of its NUV sensitivity as well as its high cadence (see also [\citenum{2014AJ....147...79S}] and [\citenum{2021arXiv211115608K}] about why studying the [N]UV emission of these type of objects is very important to understand them).  In addition, many different types of SNe will be studied but in particular the super-luminous SNe, which are SNe that can reach optical-UV luminosities that are 10-100 times more luminous than regular CCSNe. Why they are so extremely luminous is not yet clear (see the review of [\citenum{2012Sci...337..927G}] and reference thereto). To get more insight in the central powering engine, broad-band observations during the very early phase of the SN is crucial since lightcurve modelling is an important tool to diagnose super-luminous SNe (e.g., [\citenum{2017A&A...602A...9C}]). NUX can play a very important role in this because in this early phase, the NUV flux can outshine the optical one by several magnitudes (e.g., [\citenum{2017ApJ...845L...2T}]). 

\paragraph{Transiently accreting compact objects:} Accretion occurs in many different types of systems harboring compact objects, ranging from supermassive black holes in active galactic nuclei to stellar mass black holes and neutron stars in X-ray binaries, and white dwarfs in cataclysmic variables. Due to the high energies involved in the accretion process, these objects emit a large fraction (and sometimes most) of their luminosity in the, often very poorly studied, UV regime. There are many different subtypes of accreting compact objects, with the two most common ones that will produce the highest number of UV transients detected by NUX are the tidal disruption events (TDEs)\footnote{We note that we will also detect a large range of other NUV activity from a large variety of different types of accreting supermassive black holes. However, the TDEs will be the most extreme events we will detect.} and the accreting white dwarf systems. One of the defining characteristics of TDEs is their very blue colour, i.e., during the initial phases (e.g., [\citenum{2015Natur.526..542M,2014MNRAS.445.3263H}]). One of the important open questions is what actually powers the radiation emitted in TDEs and several different origins for the optical/UV emission have been proposed (for more details see the review of [\citenum{2021ARA&A..59...21G}]). Accreting white dwarfs emit most of their emission in the NUV\cite{1985ibs..book.....P,1995xrbi.nasa..331C,2008ChJAS...8..237G}, which, surprisingly, is a  poorly studied EM regime for this type of sources.  The NUV emission likely arises directly from the accretion disk due to viscous heating although the white dwarf itself might be prominently visible. The different processes give rise to different spectral energy distributions and how they evolve in time. The NUV plays an important role in this but the lack of NUV studies (both with respect to the number of sources and the sampling of these sources) inhibit our understanding.

\section{NEAR-UV EXPLORER}
\label{section:nux}

\begin{figure}[h]
    \centering 
    \includegraphics[height=7.5cm]{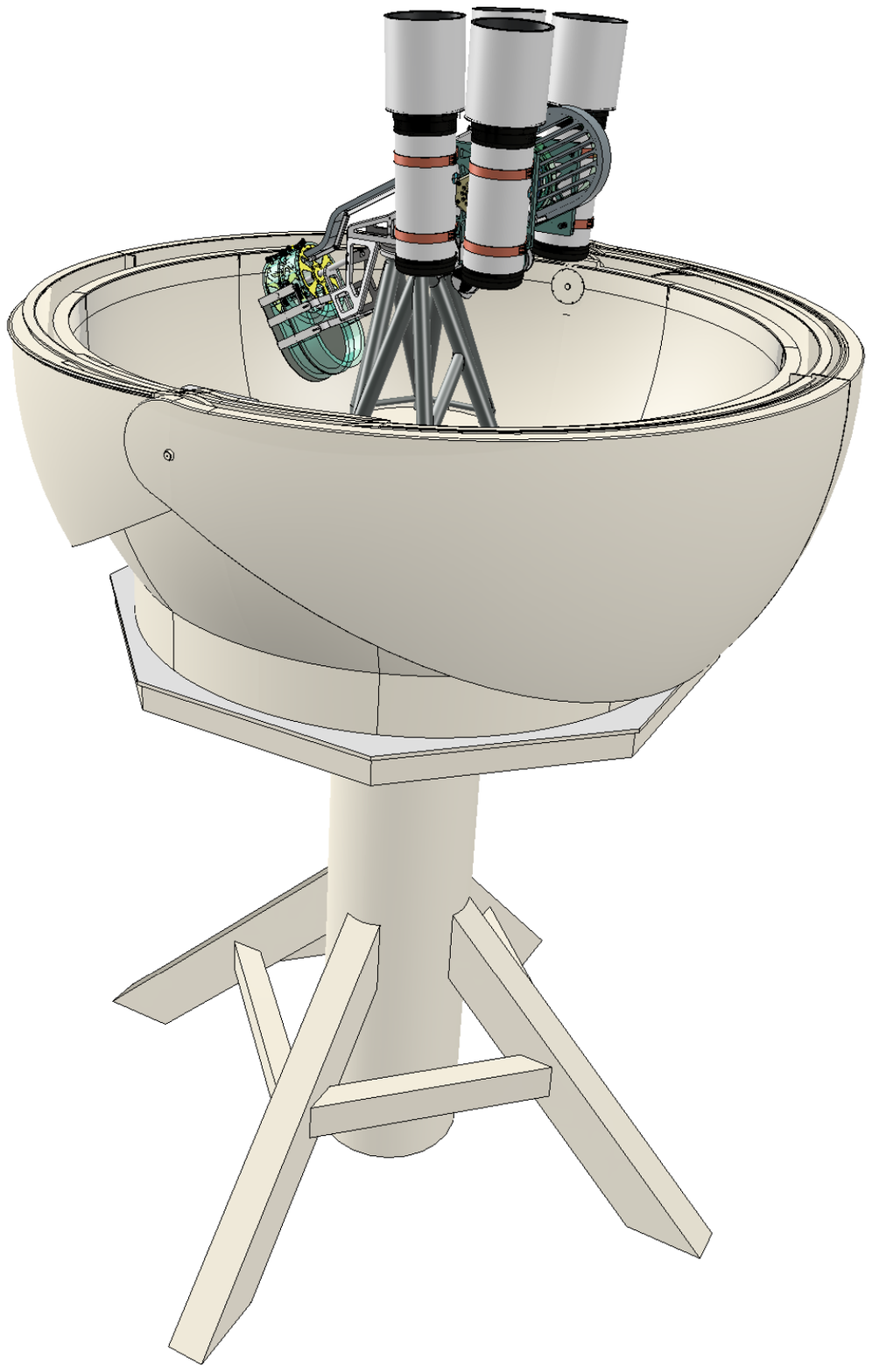}\hspace{1cm}
    \includegraphics[height=6cm]{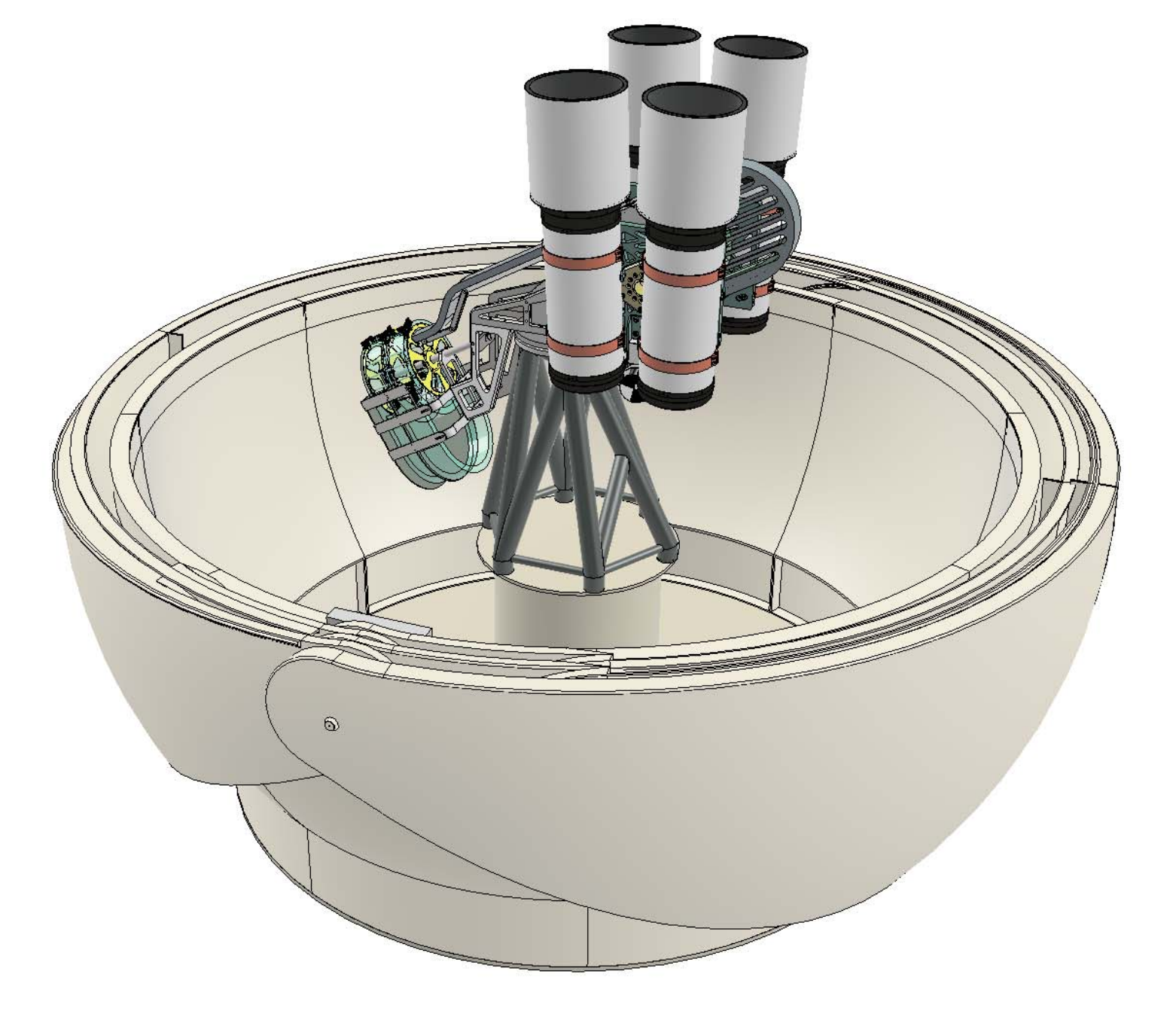}
    \caption{Computer generated view of the dome with the four telescopes on the mount in a clamshell  dome. In the left image, the impression if NUX will be located on a tower. In the right image, the impression if the tower is omitted.}
    \label{fig:NUXwithandwithouttower}
\end{figure}

\begin{figure}
    \centering
    \includegraphics[width=0.8\linewidth]{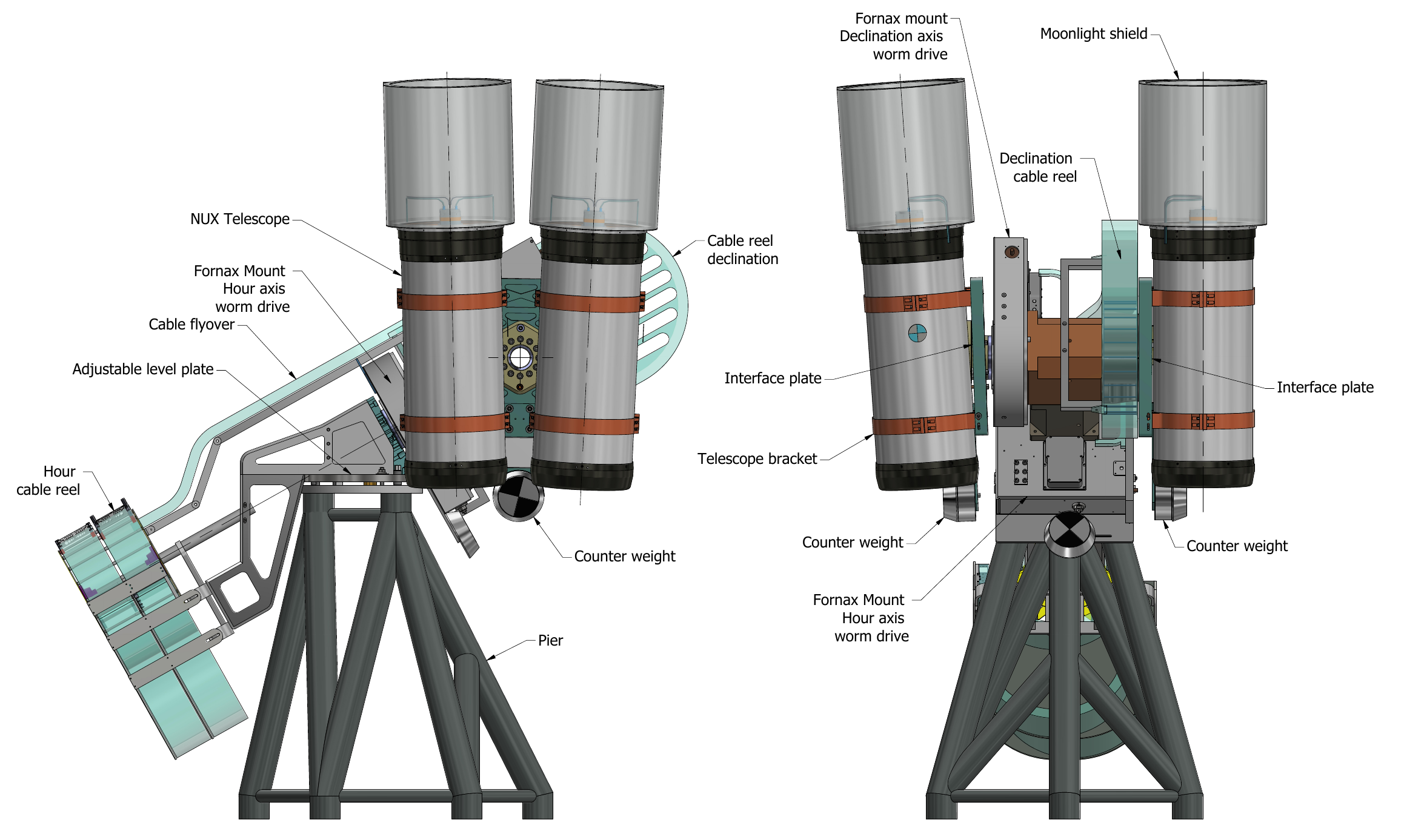}
    \caption{Computer generated image of the four NUX telescopes on top of one mount. The different components are labeled in the figures.}
    \label{fig:nuxmount}
\end{figure}

NUX is designed to image a large part of the sky at a 20-min cadence, in a cost-effective way. The design presented here uses 4 telescopes on one mount. An impression is shown in Figures \ref{fig:NUXwithandwithouttower} and \ref{fig:nuxmount}. The telescopes are adapted off-the-shelf telescopes and will have a fixed relative position on the sky, providing one large square FOV with a small overlap between the cameras. NUX will be a fully autonomous robotic facility. As a large FOV is the most important and source crowding is not an issue in the NUV, NUX will have a moderately large pixel scale of 2.6”. NUX is therefore not seeing-limited, which relaxes the optical and mechanical requirements compared to many other projects (this could mean that the tower, as presented in Fig.~\ref{fig:NUXwithandwithouttower}, left, is not needed, but this will be explored in the next phase of the project). In the following subsections, we explain the technical aspects of the project with a focus on the telescope and its optical design.

\subsection{Telescope and camera}

\begin{figure}[b]
    \centering \includegraphics[width=0.75\linewidth]{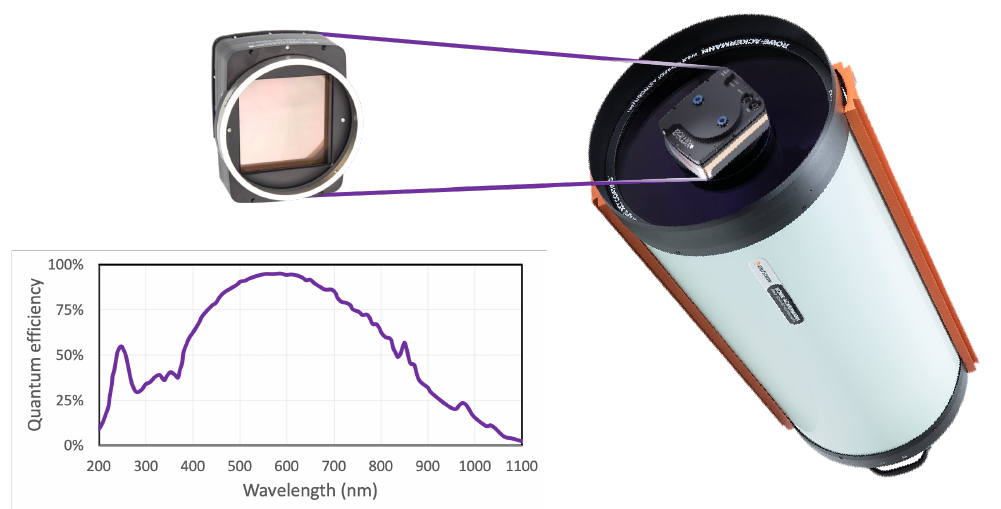}
    \caption{The 36cm RASA Celestron telescope (to the right) and the Ximea MX377MR-GP-B  camera (top left; also shown how it would fit on the telescope to the right). Below the camera photo, the quantum efficiency of the camera is given. Credits for the photos used: Celestron and Ximea. }
    \label{fig:RASAandXimea}
\end{figure}

As telescopes, we use four off-the-shelf Celestron 36cm Rowe-Ackermann Schmidt Astrograph (RASA; F/2.2) telescopes\footnote{For more details see the  \href{https://www.celestron.com/products/36-cm-rowe-ackermann-schmidt-astrograph-rasa-36-v2-optical-tube-assembly-cge-dovetail} {Celestron} web page. } (see Fig.~\ref{fig:RASAandXimea}, right) which have been optimized to have a large FOV (usable diameter of 5.1 degrees).  This makes them very suitable for large FOV surveying facilities. However, the optics in the telescopes are optimized for the visual wavelength range and therefore they have to be retrofitted with NUV transparent (the corrector and the camera lenses) and reflective (the mirror) optics. 

We have purchased one such RASA telescope to investigate the construction of the telescope as well as to create a NUV transparent corrector. It was found that the construction was sufficient for our purposes; for the manufacturing of the new NUV transparent corrector, we refer to Section \ref{section:corrector}. The NUV transparent lenses will be made from Suprasil 311/312 and CaF$_2$.

As camera, currently we focus on the Ximea MX377MR-GP-B camera\footnote{For more details see the \href{https://www.ximea.com/en/products/cameras-filtered-by-resolution-and-sensors/high-resolution-large-scmos-sensor-gpixel-gsense6060-camera}{Ximea} web page. } (see Fig.~\ref{fig:RASAandXimea}, top left) which has a Gpixel GSENSE6060BSI sensor and has a quantum efficiency in the 300-350 nm range of 35\%-40\% (Fig.~\ref{fig:RASAandXimea}, bottom left). However, as can be seen from Figure~\ref{fig:RASAandXimea} (bottom left), the detector is sensitive up to 1100 nm and thus NUV filters have to be added to the set-up so that only the desired wavelength range is detected. 

\subsection{Mount}

\begin{figure}
    \centering
    \includegraphics[height=5cm]{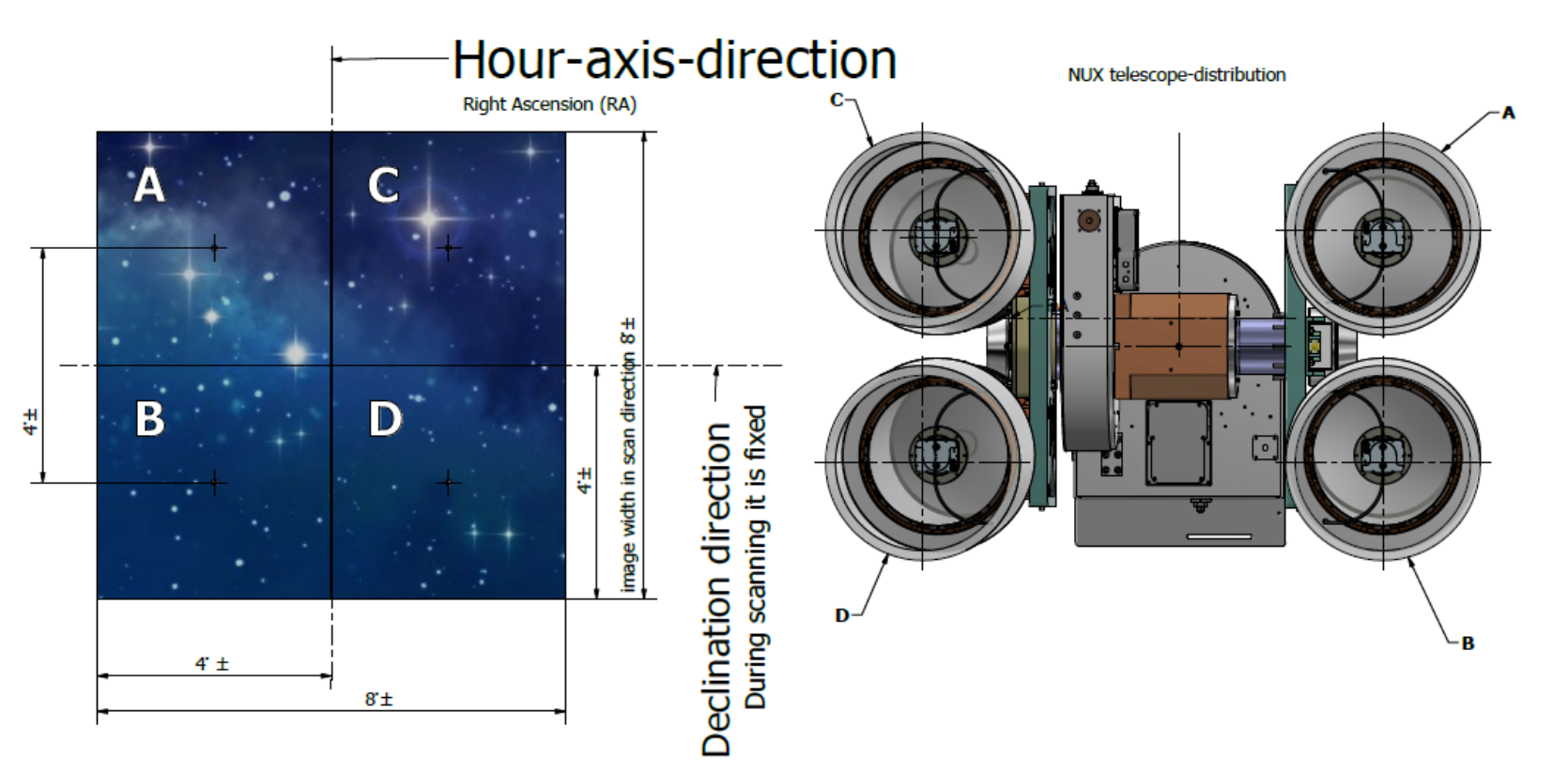}
    \hspace{1cm}
    \includegraphics[height=5cm]{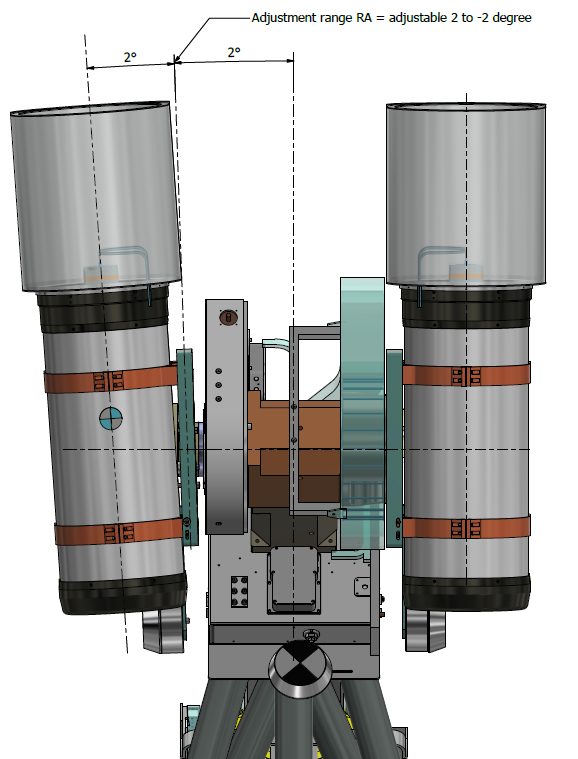}
    \caption{Left: Distribution and combined FOV of four NUX telescopes on one Fornax 200 mount. Right: Adjustment range of telescopes in right ascension.}
    \label{fig:FoVoffset}
\end{figure}

The starting point for the mount design for NUX is an existing Fornax 200 mount, similar to what is used in the BlackGEM project (see [\citenum{2016SPIE.9906E..64B,2022SPIE12182E..1VG}]). The BlackGEM mount infrastructure can be adopted almost unchanged in the NUX project. Such a Fornax 200 mount is mechanically suitable to suspend four NUX telescopes, each of which weighs around 56 kg. The two-sided suspension of the telescopes on this mount gives a favourable weight distribution. Figure \ref{fig:FoVoffset} shows the distribution of the four telescopes on the mount (see also Fig.~\ref{fig:nuxmount}) and the combined FOV achieved by offsetting the axes of the telescopes with respect to each other. The offset in right ascension is adjustable within a range of [--2.0$^\circ$, 2.0$^\circ$] by changing the orientation of the interface plate with respect to the mount. 

\subsection{NUX filters} \label{sec:filters}

For the final NUX set-up, we require a filter that has the highest transmission in the 300-350 nm wavelength range with the lower and upper wavelength boundary as close as possible to 300 nm and 350 nm, respectively. In addition, any redleak (transmission above the desired wavelength range) should be minimized and should be no more than 0.01\%. Any stronger redleak would result in a significant number of long wavelength photons to be detected and therefore would distort the inferred NUV magnitude (it would be lower than its true value; thus the detected source would appear brighter than it really is).  For the test observation using proto-NUX (see Section~\ref{section:protonux}), we also require two smaller band NUV filters, specially in the range 300-325 nm (called the "short-NUX" band) and 325-350 nm (the "long-NUX" band)

\begin{figure}
    \centering
    \includegraphics[width=0.30\linewidth]{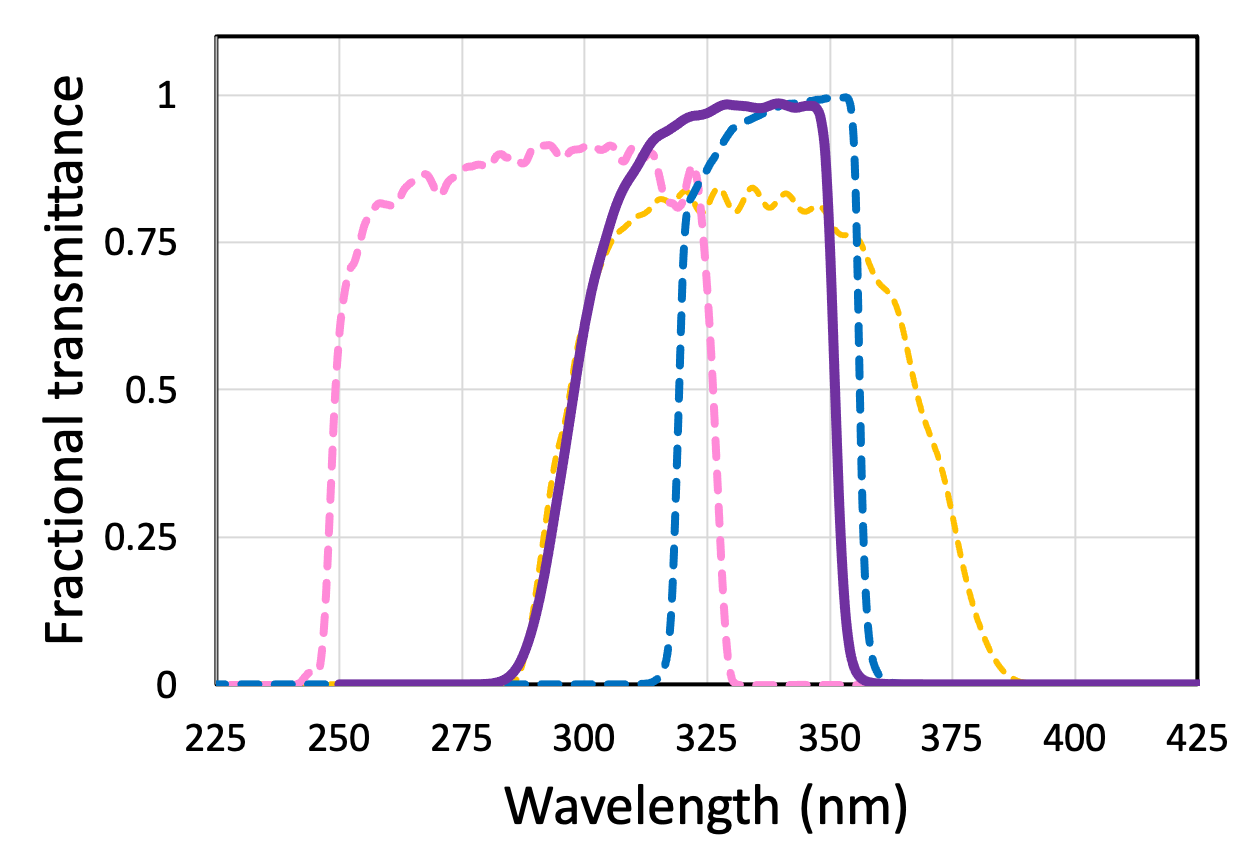}
    \includegraphics[width=0.30\linewidth]{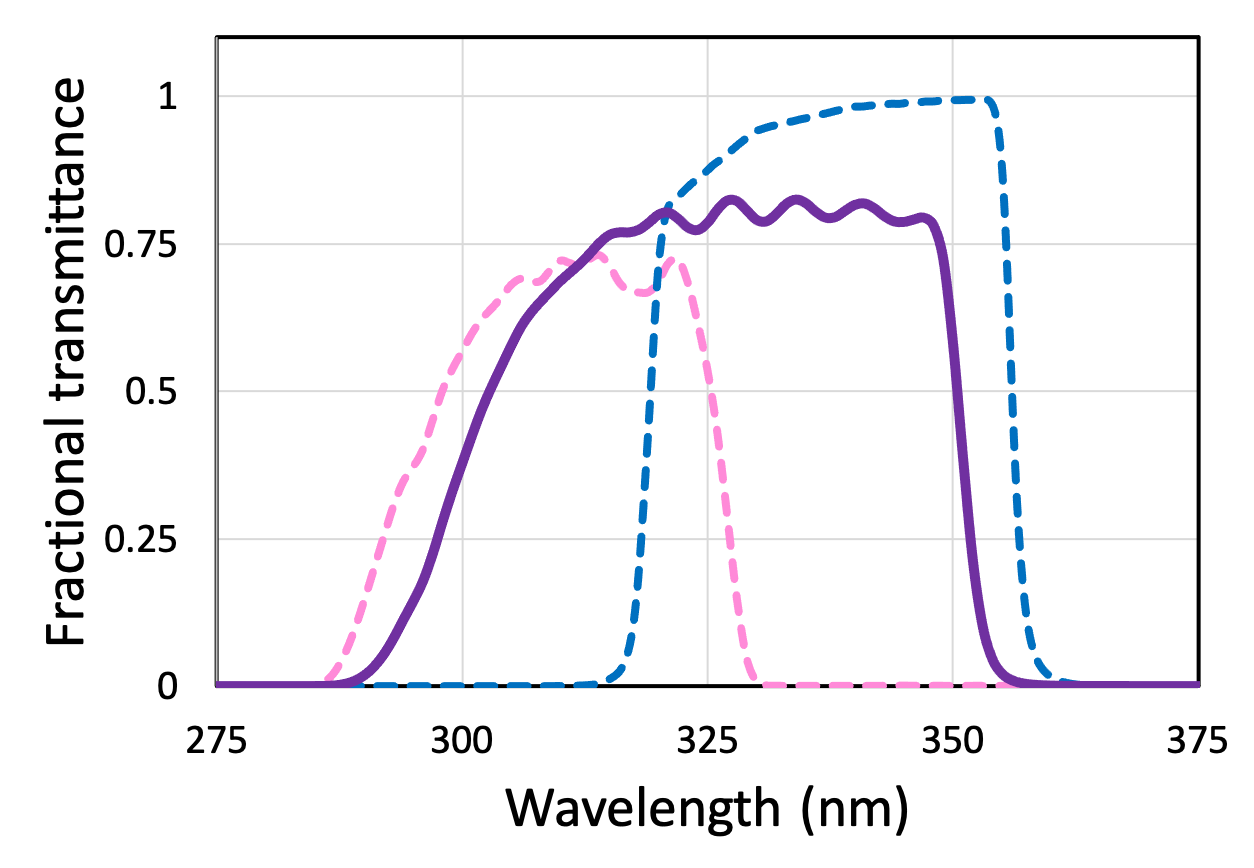}
    \includegraphics[width=0.30\linewidth]{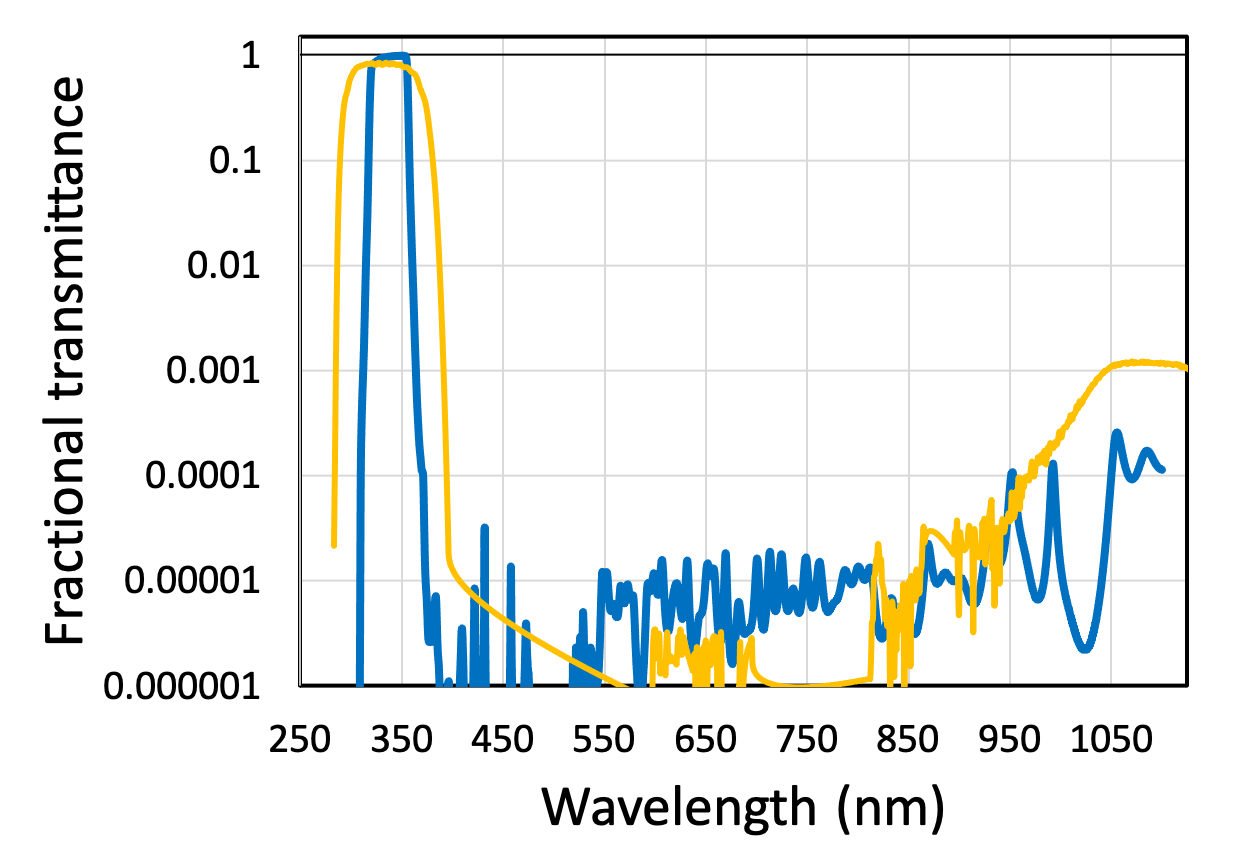}
    \caption{The current planned filters for proto-NUX. To the left, the filters as provided by Asahi: the solid purple line represent the standard NUX band (off-the-shelf), the dashed pink one represents the short-NUX band (off-the-shelf), the dashed blue one long-NUX band (custom made), and the dashed yellow line represents the red-blocking filter (off-the-shelf). We note that the short-NUX filter has significant transmittance below the desired 300 nm cut-off, but the transmission of the red-blocking filter and that of the Earth’s atmosphere below 300 nm are such that effectively this filter will have a 300 nm cut off. The middle figure shows the off-the-shelf NUX filters multiplied with the transmission of the red-blocking filters. The custom made long-NUX filter is not multiplied because of the already very low redleak of this filter. To the right, the transmittance over the full NUV-optical wavelength range showing the very low redleak for the custom made long-NUX band (blue) and the red-blocking filter (yellow).}
    \label{fig:filters}
\end{figure}

Currently, we are investigating NUV filters from several suppliers. Most readily available NUV filters have significant redleak ($>90$\% transmission at the longest wavelengths). A custom design or a separate red-blocking filter will be needed to comply with the requirements. For the test observations to be performed with proto-NUX (Section \ref{section:protonux}), we will use off-the-shelf NUV filters in combination with a red-blocking filter. The filters we are considering are listed in Table~\ref{tab:filters} and their transmission curves are shown in Fig.~\ref{fig:filters}. The red-blocking filter significantly reduces the transmittance for the NUX and the short-NUX band (the custom-made long-NUX band already has a very low redleak so the red-blocking filter will not be placed in front of it). For the test observations with proto-NUX this is acceptable because a longer exposure time than desired for NUX is not an issue. 

\begin{table}[h]
    \centering
        \caption{The Asahi filters currently under investigation for proto-NUX. }
        \vspace{0.15cm}
    \begin{tabular}{|c|c|c|}
        \hline
        Band & Asahi filter  & Impact redleak \\
           \hline
       NUX &  \href{https://www.asahi-spectra.com/opticalfilters/detail.asp?key=XHS0350}{XHS0350} & Severe \\
           \hline
      Short-NUX & \href{https://www.asahi-spectra.com/opticalfilters/detail.asp?key=XUV0325}{XUV0325}  & Severe\\
             \hline
     Long-NUX& Custom made  & Acceptable \\
            \hline
    Red-blocking & \href{https://www.asahi-spectra.com/opticalfilters/detail.asp?key=XRR0340}{XRR0340} &  Acceptable \\
            \hline
     \end{tabular} 
    \label{tab:filters}
\end{table}

\subsection{La Silla and BlackGEM} \label{sec:hosting}

The default observing site to host NUX will be the La Silla Observatory which is managed by ESO (although alternative sites are not yet fully excluded, such as the El Leoncito Astronomical Complex in Argentina). The site is well established as a good location for astronomical observatories with operations dating back to the 1960s. Facilities at La Silla, besides telescopes and control rooms, include dormitories for staff and visitors, storage areas, and workshops.  An additional benefit of La Silla is the presence of the BlackGEM array (see [\citenum{2016SPIE.9906E..64B,2022SPIE12182E..1VG}]). The co-location with BlackGEM allows for cost-savings because significant parts of the infrastructure will already be in place (network, power) which NUX can hook on to. Operational cost-savings will be possible by combining maintenance runs and spares between BlackGEM and NUX, and contacts with people on-site can also be coordinated between the two projects.

NUX uses the very bluest part of the optical window. The Chilean Andes has one of the highest UV input in the world (e.g., [\citenum{2023BAMS..104E1206C}]; see also Fig.~\ref{fig:UVChile}, left). In Fig.~\ref{fig:UVChile}, right, we show the transmission of the atmosphere in the relevant wavelength regime above La Silla (2350 m above sea level), as calculated using SKYCALC\cite{2012A&A...543A..92N} (Sky model Calculator\footnote{\url{https://www.eso.org/observing/etc/bin/gen/form?INS.MODE=swspectr+INS.NAME=SKYCALC}}) provided by ESO. This figure shows that the atmospheric extinction in the NUX 300-350 nm band is mostly caused by two different processes: Rayleigh scattering dominating the extinction above $\sim$325 nm (in the long-NUX band) and absorption by ozone for the shorter wavelengths (in the short-NUX band; Fig.~\ref{fig:UVChile}, right; ozone absorption becomes important at $\sim$325 nm and dominates below $\sim$310 nm). In addition, there is $\sim$5\% extinction due to Mie scattering by aerosol (mostly due to haze) throughout the 300-350 nm range.

\begin{figure}[h]
    \centering
    \includegraphics[width=0.7\linewidth]{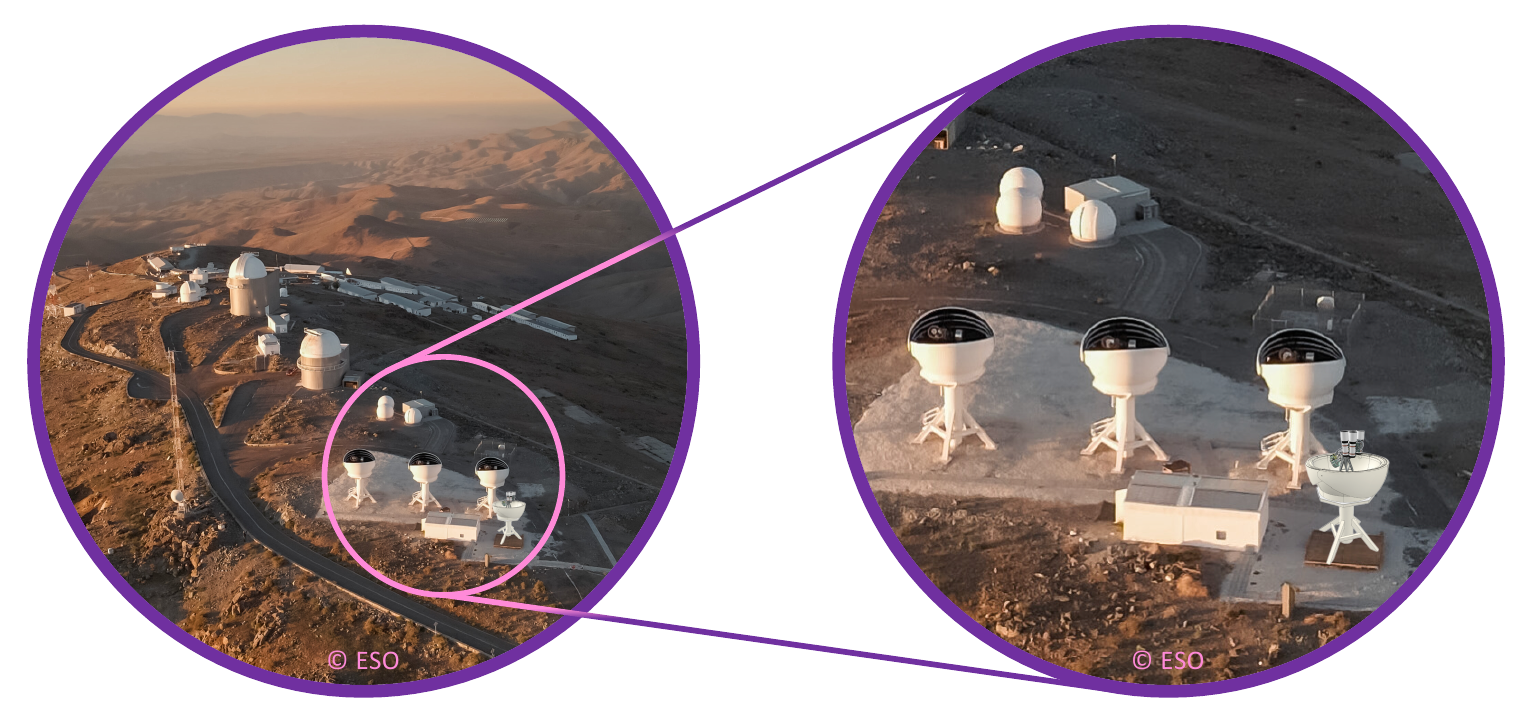}
    \caption{The NUX site on La Silla with NUX on the designated BlackGEM platform (image credits: ESO). }
    \label{fig:NUXBlackGEM}
\end{figure} 

In order to determine the sensitivity of NUX, not only did we determine the NUV throughput of the atmosphere, but we have used the SKYCALC model to calculate the night sky brightness (NSB) in the NUV. This model includes all relevant components, such as airglow, zodiacal light, and scattered moon- and starlight. The resulting sky background spectral radiance shows that the NSB in the upper NUX band ($>$325 nm) is very similar to that of the U-band, but that the NSB in the lower NUX band ($<$325 nm) drops off rapidly with wavelength. This likely is due to the fact that the spectra of the background radiation sources drop off with wavelength (see Figure 1 of [\citenum{2012A&A...543A..92N}]) as well as that it has to pass through the ozone layer which causes a steep decline of the throughput in this band. 

\subsection{Expected performance}

Our simulated optical performance of the NUX design demonstrated (using polychromatic spot diagrams) that the RMS spot radii were much smaller than the pixel size, across the NUX band and over the entire FOV. The geometric spot radii slightly exceed the pixel size. In addition, the expected vignetting is rather low, of order 10\% loss at the edge of the FOV. When incorporating the new NUV transparent optical component, the FOV of one NUX telescope will be $\sim$17 square degree, giving a total of roughly 68 square degrees (some FOV is lost because of the overlapping sky region of each telescope; see Fig.~\ref{fig:FoVoffset}). 

With the estimated transmission curve of the corrector, the primary mirror, the field corrector lenses, the detector, the NUX filter, as well as our best estimate of the transmission curve through the Earth’s atmosphere and its NUV NSB, we have determined the sensitivity of our telescope and found the limiting magnitude to be 20 mag (AB; 300-350 nm) in 2.5 minutes (dark time; at zenith). In addition, we have simulated light curves of the NUX prime targets to determine the effects of different zenith angles on the atmosphere throughput, demonstrating that we can observe our targets to a zenith angle of 45$^\circ$ without a large drop in observed brightness (less than half a magnitude) and up to 60$^\circ$ with only a 1 magnitude drop. However, these estimates are theoretical and it needs to be determined what the true decrease is with increasing zenith angle and this will be done using proto-NUX (see Section~\ref{section:protonux}).

\begin{figure}
    \centering
    \includegraphics[height=5.6cm]{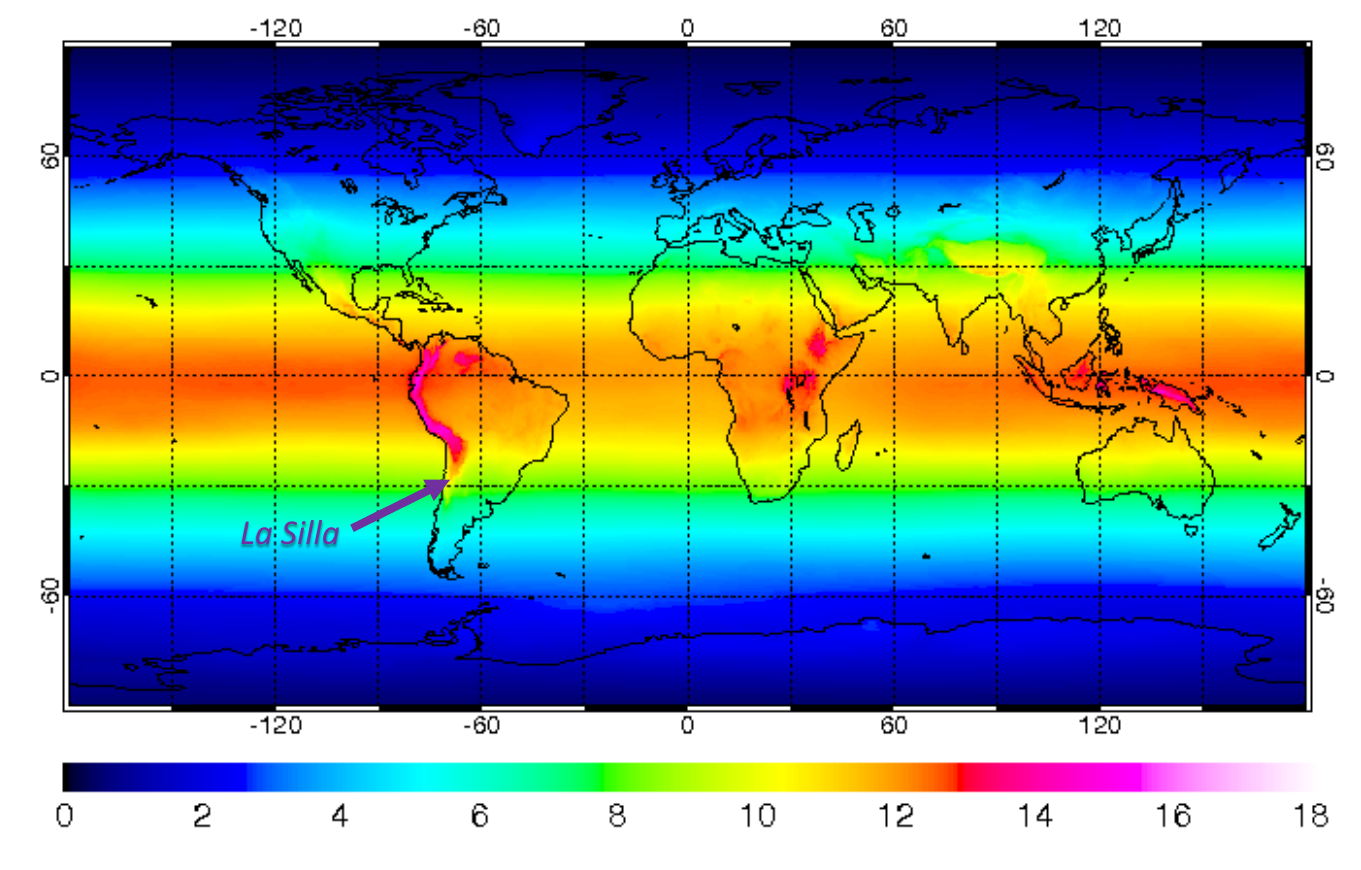}
    \hspace{1cm}
    \includegraphics[height=5.5cm]{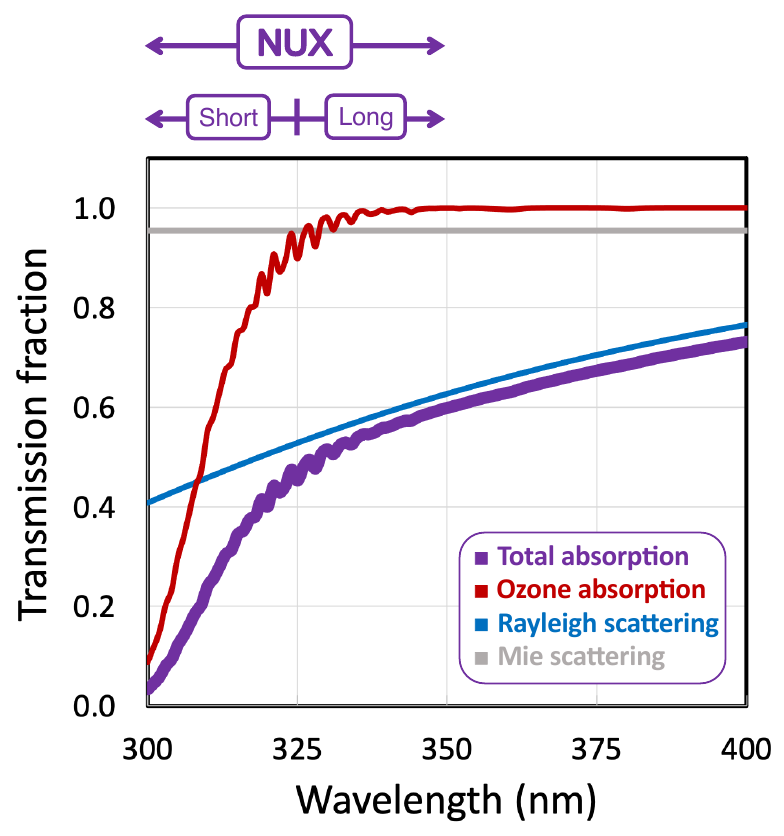}
    \caption{Left:  the UV-index at noon averaged for the years 1996 to 2002 based on data obtained using the GOME spectrometer aboard the ERS-2 satellite (credits: ESA and KNMI; obtained from \url{https://en.wikipedia.org/wiki/Ultraviolet_index}). The approximately located of La Silla is indicated in the figure with an arrow. Right: The transmission fraction in the NUV of the Earth's atmosphere. The purple line indicates the total transmission curve, while the others indicated the absorption caused by the three main components (as labelled in the figure). Above the figure, the wavelength ranges of the three NUV filters used in proto-NUX are indicated.}
    \label{fig:UVChile}
\end{figure}

\section{THE NUV TRANSPARENT CORRECTOR} \label{section:corrector}

\subsection{Polishing}

Accurate aspherical surfaces are in general not easily made. For the Schmidt corrector this is not much different. However, over the years a reliable method has been developed by co-author Rik ter Horst at the NOVA Optical/IR lab in Dwingeloo, the Netherlands, based on the manufacturing method of Bernard Schmidt, the inventor of the Schmidt system himself\cite{Everhart1966}. Here a detailed description of the method used for making this aspherical corrector is provided.

First the Suprasil blank is ground flat on both sides (within 2 micron), and made parallel within a few microns. One side is polished on the lapping machine, not necessarily flat, but at least rotationally symmetric.

For processing one aspherical surface, the pre-polished blank is placed on a vacuum pan, with the edge of the blank resting on a polished flat rim. The vacuum can be regulated with a plunger, and a spherometer is used to measure the amount of deflection, or sag. When the plunger is moved outwards, the plate is slightly pulled by the vacuum and deforms. Once the desired deflection is achieved, a sphere of specific radius is ground and polished into the deformed surface. After polishing a sphere, the sag is measured. The accuracy of the surface is then tested with a test glass; when the observed fringes are regular and straight the sag is noted. The vacuum is then released, and as the blank relaxes the outer surface will have the required aspherical shape. In this state the sag is measured once more and noted. The difference in sag, between that measured after polishing a sphere and after releasing the vacuum, is calculated to yield the total deflection, which should equal the desired deflection. In order to get a bi-aspherical corrector with a 50/50 distribution of aspherical power, the same steps are repeated for the second surface of the blank. In case the deflection of the first surface is (slightly) off, the amount of deflection of surface two can be adjusted to compensate for the deviation.

For a given set of parameters (corrector surface profile, material properties, tool parameters, etc.) the desired deflection and sag values are calculated beforehand. In the Zemax optical design, the 2nd and 4th order terms describe the deformation of the aspherical surface of the corrector. These terms and other parameters are entered in a spreadsheet tool in order to calculate the needed deflection.

\begin{figure}
    \centering
    \includegraphics[width=1\linewidth]{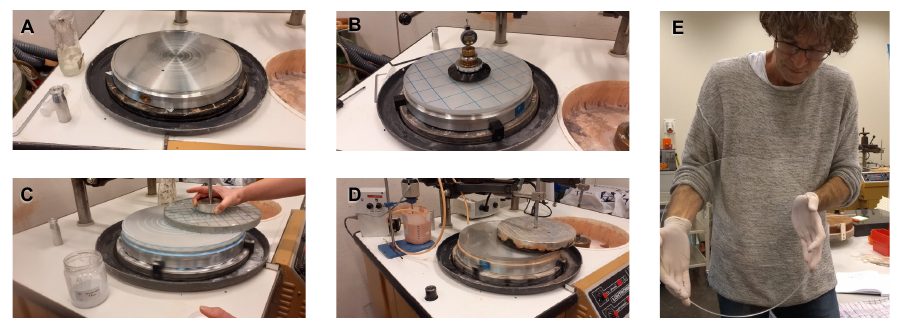}
    \caption{Photos taken during the manufacturing of the NUV transparent Schmidt corrector. A: Vacuum pan with optically polished rim. B: Measuring deflection of the corrector attached to the vacuum pan. C: Fine grinding the corrector. D: Polishing the corrector. E: The final corrector, ready for testing.}
    \label{fig:corrector}
\end{figure}

The image sequence in Figure \ref{fig:corrector} shows the steps of the manufacturing process of the Schmidt corrector. Prior to integrating the Schmidt corrector into the telescope (i.e., when assembling proto-NUX; Section \ref{section:protonux}), a 125 mm diameter hole will be drilled into the glass using a diamond core drill. The optical surfaces will be protected and a guiding tool will assure that the perforation is well centered.

\subsection{Verification}

Verification of the Schmidt corrector was performed using two different tests, one is a Ronchi test and the other an interferometric test. In both cases the Schmidt corrector is tested together with the spherical mirror of the Celestron tube to be used in NUX, but without the two field corrector lenses. As long as the Schmidt corrector is in the production and testing phase, it cannot be perforated yet, as it would be impossible to rework it using the vacuum tool. This implies also that the test system would not give a perfect image without the field lenses. To overcome this issue, a plano-convex lens was modelled to be attached onto the Schmidt corrector and using this lens we can compensate for the lack of field lenses. This configuration gives a null, that is, the total optical system under test is equivalent to a spherical mirror.

\subsubsection{Ronchi test}

\begin{figure}[t]
    \centering
    \includegraphics[width=1\linewidth]{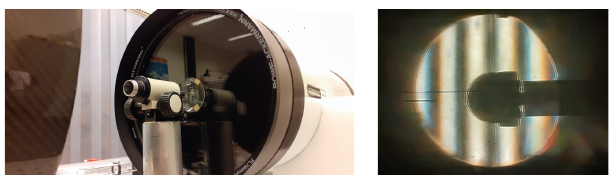}
    \caption{Left: Photo of Ronchi test setup. On the right is the Celestron tube with the Schmidt corrector under test installed. The null lens is temporarily attached to the centre of the Schmidt corrector. In front of the null lens is a microscope which creates a bright point source at the focal plane. Off the left edge is the collimator. Right: Photo of the Ronchi test result.}
    \label{fig:ronchitest}
\end{figure}

A photograph of the Ronchi test setup is shown in Figure \ref{fig:ronchitest}, left. A small, bright light source with pinhole is placed at the focal plane of the NUX system with the null-lens attached to the centre of the Schmidt corrector. From here light is reflected on the NUX primary and propagates back through the Schmidt corrector as a collimated beam towards a calibrated collimator, for which a 300 mm Newtonian telescope with accurate primary mirror (1/10 wavelength peak-to-valley wavefront error) is used. The entire optical system is tested at the focal plane of the collimator using a Foucault knife, eyepiece, and Ronchi mask.

The Ronchi grating breaks up the incident beam into multiple diffracted orders that propagate to the eyepiece and interfere to create interference patterns that are characteristic of the aberrations present in the system under test. For a perfect spherical system the resulting fringe pattern should be parallel bands of equal width. Moving the Ronchi grating along the optical axis causes the width of the bands to change. In general, the bands become narrower the farther the grating is from the zone’s centre of curvature, and conversely, the bands become wider the closer the grating is to the zone’s centre of curvature.

Figure \ref{fig:ronchitest} (right) shows a photograph of the Ronchi pattern obtained with the manufactured Schmidt corrector. The pattern exhibits straight, parallel lines as expected for a spherical system. When observing at the focal plane of the collimator, the secondary mirror of the collimator is visible as well as the spider arms that hold the secondary mirror. Residual colors are caused by the chromatic aberrations of the null lens and can easily be avoided by using a monochromatic filter or light source.

The result from this test indicates no serious deviations from the desired shape of the Schmidt corrector.

\subsubsection{Interference test}

While the Ronchi test is a relatively easy way to qualitatively check the correctness of the Schmidt corrector, it is not well suited to quantify (small) defects. To address this issue an interference test was also used.

\begin{figure}
    \centering
    \includegraphics[width=0.75\linewidth]{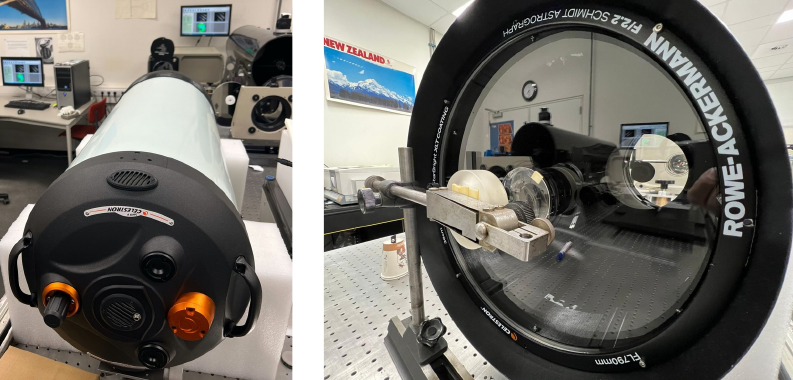}
    \caption{Photo of interference test setup. Left: As seen from behind (primary mirror side) the NUX telescope. The Wyko 6000 interferometer used in the test is visible in the background, just above the telescope. Right: The spherical mirror placed in front of the Schmidt corrector and nulling lens to reflect the beam back through the NUX optics towards the interferometer.}
    \label{fig:interferencetest}
\end{figure}

The interference test setup is shown in Figure \ref{fig:interferencetest}. A Wyko 6000 interferometer which outputs a 150 mm diameter light bundle is placed off-axis in front of the NUX telescope. The bundle is reflected from the NUX primary mirror and focused after the Schmidt corrector under test and the attached nulling lens. Behind the focus a small spherical mirror is used to reflect the light bundle back along the same path towards the interferometer. An interferogram is made and then analysed using the software Intelliwave. 

The wavefront errors and aberration parameters estimated from the interferogram are listed in Table \ref{tab:interferencetest}. The raw absolute errors as computed by Intelliwave are shown in the middle column. Since light passes twice through the Schmidt corrector under test in the setup used, the values are halved and then expressed as a fraction of a 300 nm wavelength, the shortest in the NUX band. The root mean square (RMS) wavefront error is around 2.2 \% of a wavelength, which is yields a Strehl ratio of

\begin{equation}
S = e^{{-(2\pi\omega_{RMS})^2}} = 0.981
\end{equation}

\noindent
where $\omega_{RMS}$ is the RMS wavefront error. That is a loss of just under 2 \% in sensitivity, which is considered acceptable.

Note that the interferometer light bundle, being smaller in diameter than the Schmidt corrector, only tests a sector of the Schmidt corrector. The innermost area portion of the corrector (within a diameter of 125 mm) would be removed in any case to position the field corrector and camera, so accuracy over that area is not of interest. Furthermore, based on the manufacturing process used, the surface, including errors to a large degree, is expected to be rotationally symmetric so that the results obtained here are taken as representative over the entire surface of the corrector. One noted caveat is that an off-axis test, as was done here, might be less sensitive to astigmatism than a test over the full aperture.

\begin{table}[h]
    \centering
     \caption{Results from the interference test. The leftmost and middle columns show the parameters computed by Intelliwave from the interferogram. The rightmost column expresses half the values as a fraction of the shortest wavelength in the NUX band; the factor of a half compensates for the double pass through the Schmidt corrector in the used setup.}   
     \vspace{0.15cm}
     \begin{tabular}{|l|r|c|}
        \hline
        Parameter & Value & Half value \\
         &  (nm)  & (in wavelengths at 300 nm)\\
          \hline
        Peak-to-valley &  150.262 & 0.215 \\
             \hline
     RMS &  15.363 & 0.022 \\
              \hline
    Focus & 8.992 &  0.013\\
          \hline
        Astigmatism & -21.465 & -0.031 \\
           \hline
       XY astigmatism & 10.374 & 0.015 \\
           \hline
       X coma &  -4.248 & -0.006\\
          \hline
        Y coma &  -7.062 & -0.010 \\
          \hline
        Spherical & -2.046  &  -0.003\\
         \hline
    \end{tabular}
    \label{tab:interferencetest}
\end{table}

\section{NUX OBSERVING STRATEGIES}
\label{section:strategies}

NUX will execute the survey described in Section \ref{section:survey}, but can also be triggered by an external event such as a GW event as described in Section \ref{section:trigger}.

\subsection{NUX survey strategy}\label{section:survey}

Due to the significant increase in atmospheric extinction with increasing zenith angle (thus increasing airmass), we plan to restrict NUX pointings to a sky patch within 30 degrees of zenith, with two exceptions: the Magellanic Clouds, as well as to follow interesting transients for a longer time. This band in the sky covers a major fraction of nearby (within 100 Mpc) galaxies and galaxy clusters in the Southern skies, including all of the Fornax Cluster, the Galactic Center and Bulge, a major part of the Galactic Plane, and the M83/M104 Southern extension of the Virgo Cluster, as well as the Centaurus A and Sculptor galaxy groups. We note that for telescopes with a FOV this large even the largest galaxies on the sky, except the Galaxy, fully fit within one frame.

Assuming a 4 telescope NUX system, having a total field of view of $\sim$68 square degrees and 2.5 minutes exposure times (to reach 20 magnitude in dark time), NUX will be able to cover about 1350 square degrees per hour (also taking into account slight overlap between observing fields and the FOV of each telescope, as well as a 10 s overhead between pointings). As NUX is aiming for high cadence observations, we will use a 20 minute cadence between observations. This means that 1350/3 = 450 square degrees per 20 minutes can be observed. An object in the maximum viewing zone, such as the Galactic Bulge and the Fornax cluster, will be visible for $\sim$4 hrs per night in the zenith patch and therefore be observed 12 x per night. In total, at least 1000 square degrees on the sky can be covered per 10-hr night when it is new Moon. This will drop by a factor of two for grey time in order to reach the desired sensitivity of 20 mag. 

\subsection{NUX trigger mode}\label{section:trigger}

Besides executing the standard survey program described in Section \ref{section:survey}, NUX can be triggered to observe specific fields. The most important triggered observations will be the follow-up of GW events detected with the GW facilities. Follow-up of such triggers will be done in an automated way, where the observatory control software will autonomously follow-up on triggers sent out by instruments such as LIGO and Virgo. In addition, any other target found by other facilities that are deemed of very high interest (e.g., an unusual SN or TDE) will be observed by NUX. However, this will require manual intervention.

\section{COMPARISON WITH ULTRASAT AND UVEX} \label{section:ultrasatuvex}

Although NUX will be a ground-breaking new transient search facility, two other initiatives are currently pursued to find and study UV transients. For example, the UV satellite ULTRASAT\cite{2014AJ....147...79S,2024ApJ...964...74S} is currently being constructed with a scheduled launch at the end of 2027\footnote{See \url{https://www.weizmann.ac.il/ultrasat/} for the most recent information about ULTRASAT.}. Although the cadence of ULTRASAT (200 square degree every 5 minutes) is very similar to that of NUX (450 square degree every 20 minutes), ULTRASAT is sensitive to shorter wavelengths (i.e., 230-290 nm) than NUX (300-350 nm) and therefore both facilities are very complementary to each other with their own specific source types that they are optimally suitable for to find and study.  

Another relevant initiative is the UVEX mission\cite{2021arXiv211115608K} which was very recently approved by NASA and is aimed to be launched in 2030\footnote{See \url{https://www.uvex.caltech.edu/} for the most recent information about UVEX.}. Its imager will be sensitive in the 203-270 nm range (as well as in the 139-190 nm range). However, UVEX is quite different that NUX and ULTRASAT in cadence because the prime goal of its imager will be to perform a synoptic survey of the entire sky with a relatively low number of pointings per sky position (a minimal of 10 times in the 2 year baseline mission, with a cadence of $\sim$12 hrs to months). Therefore, in a single day UVEX will not detect the same large number of UV transients as NUX and ULTRASAT will find, it will not obtain early time data of the transients ($<$1 hr), and it will mostly obtain detailed light curves of targeted transients (e.g., found by other facilities) and not for serendipitous sources.  However, UVEX has excellent UV spectroscopic follow-up capabilities and therefore will be very important to obtain UV spectra of the most important transients found by NUX and ULTRASAT. 

The current total expected cost of NUX will be 1.7 - 2 million EUR. ULTRASAT is budgeted at 100 million USD and UVEX even at 300 million USD. This demonstrates that NUX will be able to perform ground-based NUV transient searches at only 1\%-2\% of the costs of satellite based instruments. 

\section{PROTO-NUX}
\label{section:protonux}

During the remainder of 2024, we plan to build a complete NUX prototype telescope (called proto-NUX) using the already available Celestron telescope and the manufactured NUV transparent Schmidt corrector. This telescope will then be tested on sky in Dwingeloo to achieve a first verification of optical performance, on a Dobsonian mount. Subsequently the telescope will be shipped for further testing to La Silla, where it will be installed on a portable mount, outside (on a tripod) or in an existing dome (potentially on the available pier or, if absent, on a tripod), and where it will be operated in a hands-on fashion. Full robotic operations as intended for the final version of NUX is outside the scope of proto-NUX.  The main goals of proto-NUX are to 

\begin{itemize}
    \item demonstrate a ground-based NUV telescope can indeed be constructed successfully and that technically it performs according to expectations.
    \item characterise the throughput and the NSB of the Earth's atmosphere for the different NUX bands, both of which are poorly studied so far. This will be crucial to determine the true sensitivity of NUX and how it varies with airmass and in time (intra- and inter-night, as well as seasonal).

\end{itemize}

During the assembly of proto-NUX, we will include the new NUV transparent corrector as well as NUV transparent lenses. In addition, the NUV reflectivity of the mirror will be investigated and if found it is not sufficient, it will have to be recoated with a NUV reflective one. Also, a filter slider will be included to house the different filters. With using the three NUV bands in our prototype, we will be able to determine the (in)stability of the NUV transmittance of the Earth’s atmosphere (as well as the NUV NSB) for the first time for a variety of time scales (intra- as well as inter-night; seasonal effects), and determine if the $<$325 nm extinction significantly deviate from that observed for wavelengths of $>$325 nm. Uncorrelated behavior in both bands might be possible because of the different dominate absorption processes in these bands (ozone absorption vs. Rayleigh scattering, respectively; Fig.~\ref{fig:UVChile}, right). The outcome of these investigations will be very important for optimising the NUX observing program as well as for accurately calibrating the observed light curves of our targets.

\paragraph{Calibration:} In preparation of the tests in Chile, we will compose a photometric calibration catalogue for the NUX bands, essential for accurate sky-transparency tests. This catalogue will be based on the low-resolution BP/RP spectra available in Gaia DR3 for 220 million of stars\cite{2023A&A...674A...2D}. These spectra cover the wavelength range 330-1050 nm, and so are not directly applicable to the NUX passband. However, the Gaia spectra can be fitted with model spectra from stellar libraries\cite{2003IAUS..210P.A20C,1985ApJS...59...33P}, including Galactic extinction, to infer the most likely spectrum across the NUX bands. On the blue side, these spectral fits can be constrained by GALEX observations for a number of stars. Also, a subset of blue stars will be selected to ensure sufficient flux in the NUX passband. Once the best-fitting spectrum has been inferred, the magnitude of the calibration star in the proto-NUX system can be determined, taking into account the estimated typical atmospheric extinction in the NUX filters and the transmission of the different components in proto-NUX itself, such as the mirror reflectivity, filter transmissions, and detector quantum efficiency. With this NUX calibration catalogue, each image can be directly calibrated using at least 100 calibration stars in the field. A similar strategy will be employed for the final NUX facility and performed for each of the four telescopes.

\paragraph{After proto-NUX:} Currently, proto-NUX is fully funded by the University of Amsterdam. If proto-NUX and the NUV characteristics of the Earth's atmosphere are as expected and thus we have demonstrated that a ground-based NUV telescope is feasible and viable, future funding for NUX needs to be obtained. Several funding schemes will be explored (e.g., using the funding options provided by the European Research Council) but the NUX project is open for additional partners. If interested, please contact the first author of this paper.

\acknowledgments 
 
NUX has received funding from the Netherlands Research School for Astronomy (NOVA) and proto-NUX from the University of Amsterdam.

\newpage
\bibliography{report} 
\bibliographystyle{spiebib} 

\end{document}